\documentclass[aps,twocolumn,superscriptaddress,a4paper]{revtex4}
\usepackage{graphicx}

\newcommand{\beq}{\begin{equation}}
\newcommand{\eeq}{\end{equation}}
\newcommand{\beqa}{\begin{eqnarray}}
\newcommand{\eeqa}{\end{eqnarray}}

\newcommand{\bigmean}[1]{\left\langle#1\right\rangle}
\newcommand{\dd}{{\rm d}}
\newcommand{\del}{\delta}
\newcommand{\e}{{\rm e}}
\newcommand{\ebar}{{\overline{\varepsilon}}}
\newcommand{\eff}{{\rm eff}}
\newcommand{\eps}{\varepsilon}
\newcommand{\hi}{{\rm high}}
\renewcommand{\max}{{\rm max}}
\newcommand{\mf}{{\rm MFT}}
\newcommand{\mean}[1]{\langle#1\rangle}

\newcommand{\prob}{\mathop{\rm Prob}\nolimits}
\newcommand{\scr}{{\rm SCR}}
\newcommand{\ty}{{\rm typ}}
\newcommand{\Del}{\Delta}
\newcommand{\KK}{\widetilde{K}}

\begin{document}
\title{How the fittest compete for leadership: A tale of tails}

\author{J.~M.~Luck}\email{jean-marc.luck@ipht.fr}
\affiliation{Institut de Physique Th\'eorique, Universit\'e Paris-Saclay, CEA and CNRS,
91191 Gif-sur-Yvette, France}

\author{A.~Mehta}\email{anita.mehta@ipht.fr}
\affiliation{Institut de Physique Th\'eorique, Universit\'e Paris-Saclay, CEA and CNRS,
91191 Gif-sur-Yvette, France}
\affiliation{Dipartimento di Fisica,
Universit\`a di Roma La Sapienza, P.~A.~Moro 2, 00185 Roma, Italy}
\affiliation{Institut f\"ur Informatik,
Universit\"at Leipzig, H\"artelstrasse 16-18, 04107 Leipzig, Germany}

\begin{abstract}
We investigate how leaders emerge as a consequence of the competitive dynamics
between coupled papers in a model citation network.
Every paper is allocated an initial fitness depending on its intrinsic quality.
Its fitness then evolves dynamically as a consequence of the competition
between itself and all the other papers in the field.
It picks up citations as a result of this adaptive dynamics,
becoming a leader if it has the highest citation count at a given time.
Extensive analytical and numerical investigations of this model suggest the
existence of a universal phase diagram,
divided into regions of weak and strong coupling.
In the former, we find an `extended' and rather structureless distribution
of citation counts among many fit papers;
leaders are not necessarily those with the maximal fitness at any given time.
By contrast, the strong-coupling region is characterised by a
strongly hierarchical distribution of citation counts,
that are `localised' among only a few extremely fit papers,
and exhibit strong history-to-history fluctuations,
as a result of the complex dynamics among papers in the tail of the fitness distribution.
\end{abstract}

\maketitle

\section{Introduction}
\label{intro}

The field of complexity has gained greatly in importance in recent times, in
part because examples of such systems abound in the real world, and in part
because advances in numerical and analytical techniques enable their detailed examination.
Physical systems such as earthquakes and sandpiles, social systems
such as communities, financial systems such as stock markets, and biological
systems such as the human brain all manifest complexity~\cite{gellmann}.
These are all examples
of systems whose many components interact with each other dynamically, leading
to the emergence of collective effects that are non-trivial and often unexpected.
Typically, these interactions are non-linear,
which is a key reason behind some of the surprising outcomes.
Irreversibility and history-dependence are other key ingredients of such
systems, which are typically far from equilibrium.

Statistical physics has usually concerned itself with trying to model real,
complex phenomena
by using a variety of tools, of which one of the most important is agent-based
modelling~\cite{bonabeau}.
Here, agents on a lattice or other network interact according to the domain
under consideration, be this traders in a stock market, or genes in a gene network.
Our own work in this domain has ranged from
black hole accretion~\cite{usar} to more abstract examinations of competitive
dynamics~\cite{us}.
The present paper is the culmination of a body of work starting from the
latter, where the following question was raised:
who are the survivors in a given scenario of competitive or predatory dynamics,
and what determines their survival? Our findings
were that the `survival of the fittest' is not always a given in such a situation;
often it is the less fit who survive, in a situation we have referred to as
`winning against the odds'~\cite{nirmal}.
The addition of spatial complexity,
via a network with random nodal connectivities, provides ways for
outliers to hide from, and sometimes survive, the overt competition of hubs~\cite{suman}.
More recently, we were able to identify universality in the statistics of survivors
among competing agents on networks~\cite{usrec}.
These findings were reminiscent of the universality found in various studies,
empirical as well as theoretical,
of citation networks, which focused on the citation counts of single
papers~\cite{rev4,rev3,wsb,kp,mp,ref1,xo,zz,ref3,ref2}.

Our motivation for the present work is to understand how such universality
might come about,
which has led us to propose a model in the context of citation networks.
Our emphasis is, however, on
collective rather than individual dynamics: thus, rather than focusing on the
citation counts of a single paper, we examine that of an ensemble of papers in a
specific discipline, each one with a given initial fitness.
These papers, as in
real life, are coupled both to their predecessors and their successors, which
leads to a dynamical evolution of their fitness.
The strength of the coupling constant $g$
is crucial to the adaptive dynamics that characterise this evolution.
The results of our analytical and numerical work will demonstrate that when papers
are weakly coupled, the citation counts they acquire during their lifetimes are
well described by mean-field dynamics.
In the limit of strong coupling, on the
other hand, we will see that a few very fit papers have the lion's share of
citation counts, and simple mean-field theories are no longer adequate to
describe them.
The competitive dynamics that occur in the tail of the fitness
distribution give rise to phenomena which can justifiably be called complex, of
which a striking example is the fact that the paper that has the highest
citation counts (the so-called leader or `winner') at any given time is not necessarily
the one with the highest fitness (the so-called `record').
It is this competition among the fittest papers which
is both the most novel and the most important ingredient of the present study.

The model is defined in Section~\ref{model}.
The mean-field approach of Section~\ref{mft}
provides an analytical description of its steady state,
and predicts a universal phase diagram,
with a weak-coupling regime (WCR) and a strong-coupling regime (SCR)
separated by a sharp crossover near the critical coupling~$g_c$.
Section~\ref{num} contains numerical results on many quantities of interest including
the total activity, the fates of single papers and the distribution of citation
counts; in particular we examine two sequences of exceptional papers,
records and leaders, based on their fitnesses and citation counts.
These lead naturally to a discussion of the strong-coupling regime of the model,
for which we develop an effective model based on the statistics of records
in Section~\ref{effective}.
Finally, we discuss our results in Section~\ref{disc}, relegating to the
Appendix~\ref{app} a mean-field analysis of the model
for an arbitrary distribution of initial fitnesses.

\section{The model}
\label{model}

The main criteria behind the formulation of the present model are
simplicity and minimalism, i.e., we choose the least complex model that still
manages to capture the essence of fitness and citation dynamics.

A new field of study is established at some initial time ($t=0$),
as papers begin to appear in it;
we assume additionally that they appear at random times,
with a constant rate $\omega$, and are numbered in the order in which they appear.
We mention in passing that this situation can easily be generalised to one
where new papers
in the field draw on, and then compete with, papers from established fields;
this can be done by a simple modification
of the empty initial configuration here presented to an appropriately
structured, non-empty one.

Initial fitnesses $\eta_i(t_i)=\eps_i$ are allocated to papers $i$,
published at times $t_i$.
These are quenched random variables,
drawn from some probability distribution $\rho(\eps)$ with a bounded support,
i.e., $\eps_\max$ finite.
Initial fitnesses provide measures of the intrinsic quality of the papers
with which they are associated.
In this work, they are drawn from the uniform distribution on [0, 1].
A generalisation of our mean-field analysis to arbitrary
fitness distributions on [0, 1] is presented in the Appendix~\ref{app}.

In the following, the evolution of the dynamic fitness $\eta_i(t)$
of paper $i$ from its initial value $\eps_i$
is largely determined by the competition between itself and all the other
papers in the field.
We choose to model this evolution as follows:
\beq
\frac{\dd}{\dd t}\eta_i(t)=-(1+\del-\eps_i)\eta_i(t)+\gamma_i(t).
\label{eqeta}
\eeq
The first term represents the spontaneous decay in the course of time
of the fitness of a single paper in the absence of interactions.
The associated relaxation time
\beq
\tau_i=\frac{1}{1+\del-\eps_i}
\label{taui}
\eeq
is an increasing function of the initial fitness $\eps_i$,
so that fitter papers have a longer-lasting impact.
Perfectly fit papers ($\eps_i=1$) have the largest relaxation time $\tau_i=1/\del$.
The damping rate $\del$ plays the role of a regulator.

Intuitively, there are a couple of features to be taken into account when modelling
the interaction term $\gamma_i(t)$,
representing competition between papers:
\begin{enumerate}
\item the competition should be tougher for the fitter papers;
\item the intrinsic quality of papers, as measured by their initial fitnesses,
should also have a lasting effect.
\end{enumerate}

The following simple form for the interaction term is accordingly chosen:
\beq
\gamma_i(t)=g\sum_j(\eps_i-\eps_j)\eta_i(t)\eta_j(t),
\label{inter}
\eeq
where the sum runs over papers $j$ which compete with paper $i$ at time $t$,
i.e., all papers published before time~$t$,
and~$g$ is a positive coupling constant.
The requirement 1 is modelled by taking the interaction
to be proportional to the product $\eta_i(t)\eta_j(t)$
of the instantaneous fitnesses of both competitors,
and the requirement 2 is taken into account via the bias factor $\eps_i-\eps_j$.

We suggest also that papers accumulate citations stochastically, so that
any paper~$k$ quotes any earlier paper $i$ with probability $p_{k,i}$.
This citation probability is entirely dictated by the dynamic fitness~$\eta_i(t_k)$
of paper $i$ at the time $t_k$ when paper $k$ was published.
For definiteness we assume a linear law of the form
\beq
p_{k,i}=\lambda\eta_i(t_k),
\label{pki}
\eeq
where $\lambda$ is a small positive constant.

The mean number of references of paper $k$ is computed by evaluating
an average over the stochastic citation process.
This reads
\beq
R_k=\lambda\sum_i\eta_i(t_k),
\label{rex}
\eeq
where the sum runs over papers $i$ published before time~$t_k$.
The mean citation count $C_i(t)$ of paper~$i$ at time~$t$ is, analogously, given by
\beq
C_i(t)=\lambda\sum_k\eta_i(t_k),
\label{cmean}
\eeq
where the sum runs over papers $k$ published between~$t_i$ and $t$.
In particular, the mean citation count accumulated by paper $i$
during its whole history reads
\beq
C_i^\infty=\lambda\sum_k\eta_i(t_k),
\eeq
where the sum runs over papers $k$ published after $t_i$.

\section{Mean-field theory}
\label{mft}

In this section we present an approximate analytical description of the model,
using mean-field theory.
It turns out that mean-field predictions are exact for some global quantities.
Additionally, the predictions for all but the fittest individual papers are
essentially correct (see Section~\ref{num}).

\subsection{The fate of an individual paper}
\label{mftsingle}

The key idea of mean-field theory is to look at the evolution of
an entity in a mean environment, whose characteristics are then obtained self-consistently.
In this case, we look at the evolution of a selected individual paper
with given initial fitness, competing with all the others.

The subsequent analysis is limited to the steady state of the model,
when the field has matured.
The existence and the uniqueness of the steady state are guaranteed
by the finiteness of all the relaxation times~(\ref{taui}),
which in turn relies on the presence of a non-zero damping rate~$\del$.
Since the steady state of the model is invariant under time translation,
it can be assumed that the selected paper is published at time $t=0$
with no loss of generality.
This paper is characterised by its initial fitness~$\eps$, with the
subsequent evolution of its dynamic fitness $\eta(t;\eps)$ being described by
the stationary form of~(\ref{eqeta}), i.e.,
\beq
\frac{\dd}{\dd t}\eta(t;\eps)=-(1+\del-\eps-g(A\eps-B))\eta(t;\eps).
\label{mf}
\eeq
The two mean fields acting on the selected paper,
\beq
A=\Bigl\langle\sum_i\eta_i(t)\Bigr\rangle,\qquad
B=\Bigl\langle\sum_i\eps_i\eta_i(t)\Bigr\rangle,
\label{abdef}
\eeq
are independent of time $t$, since we are dealing with a steady state.
The sums in the above expressions
run over papers $i$ published before the selected paper (i.e., at negative times).
Here and throughout the following,
brackets denote averages over the whole stochastic history of the model.
In the present mean-field context,
this amounts to averaging over the fitnesses and publication times of all
papers entering the sums.

It is useful to introduce the combinations
\beq
L=1+\del+gB,\qquad M=1+gA,
\label{ablm}
\eeq
so that~(\ref{mf}) reads
\beq
\frac{\dd}{\dd t}\eta(t;\eps)=-(L-M\eps)\eta(t;\eps).
\label{mf2}
\eeq

The dynamic fitness of the selected paper then reads
\beq
\eta(t;\eps)=\eps\,\e^{-(L-M\eps)t}.
\label{mfeta}
\eeq
This exponential relaxation law for the dynamic fitness
is a key result of the mean-field approach.
The relaxation rates $L$ and $M$,
related to the mean fields $A$ and~$B$ through~(\ref{ablm}),
have a non-trivial dependence on the model parameters $\del$, $g$ and~$\omega$.

The mean number of references of a paper in the steady state is obtained
by averaging~(\ref{rex})
over the fitnesses and publication times of all other papers.
This reads
\beq
R=\lambda A.
\eeq
The mean number of references of the selected paper is an indication of the
activity
of the field, so we will, from now on, refer to $A$ as the mean activity of the
model.

The mean citation count of the selected paper at time~$t$ can be computed similarly:
\beqa
C(t;\eps)&=&\lambda\omega\int_0^t\eta(t';\eps)\,\dd t'
\nonumber\\
&=&\frac{\lambda\omega\eps}{L-M\eps}\left(1-\e^{-(L-M\eps)t}\right).
\label{mfc}
\eeqa
In particular, the mean citation count accumulated by the paper
during its whole history is predicted to be
\beq
C^\infty(\eps)=\frac{\lambda\omega\eps}{L-M\eps}.
\label{mfcinf}
\eeq
For a perfectly fit paper ($\eps=1$), this reads
\beq
C^\hi=\frac{\lambda\omega}{L-M}.
\eeq

\subsection{Mean-field equations and their solution}
\label{mftsol}

Here, we evaluate the mean fields $A$ and $B$
as well as the relaxation rates $L$ and $M$.

The mean fields obey the self-consistency equations
\beqa
A&=&\omega\int_0^1\dd\eps\int_0^\infty\eta(t;\eps)\,\dd t,
\label{asc}
\\
B&=&\omega\int_0^1\eps\,\dd\eps\int_0^\infty\eta(t;\eps)\,\dd t.
\label{bsc}
\eeqa
These equations are derived from~(\ref{abdef})
by approximating the sum over $i$ by integrals over $t=-t_i$,
the age of paper~$i$ at time $t=0$, i.e., when the selected paper appears.
Moreover, $\eta(t;\eps)$ is given by~(\ref{mfeta}).
The resulting equations can be solved by using
\beq
x=\ln\frac{L}{L-M}
\label{xmf}
\eeq
as a parameter.
Introducing the notation
\beq
\Del=2(\e^x-1)((x-1)\e^x+1)\del+\e^{2x}-2x\e^x-1,
\label{dmf}
\eeq
we obtain after some algebra:
\beqa
g\omega&=&\frac{2(\e^x\del-\del-1)\Del}{(\e^x-1)^3},
\label{gmf}
\\
L&=&\frac{\e^x\Del}{(\e^x-1)^3},
\label{lmf}
\\
M&=&\frac{\Del}{(\e^x-1)^2},
\label{mmf}
\\
A&=&\omega\,\frac{(\e^x-1)((x-1)\e^x+1)}{\Del},
\label{acmf}
\\
B&=&\omega\,\frac{(2x-3)\e^{2x}+4\e^x-1}{2\Del},
\label{bmf}
\\
C^\hi&=&\lambda\omega\,\frac{(\e^x-1)^3}{\Del}.
\label{cmf}
\eeqa

Mean-field theory becomes exact in the absence of interactions ($g=0$).
We have then $L=1+\del$ and $M=1$ (see~(\ref{ablm})).
The parameter $x$, the mean activity $A$ and the highest citation count
$C^\hi$ take their minimal values
\beqa
x_0&=&\ln\frac{1+\del}{\del},
\\
A_0&=&\omega\left(\!(1+\del)\ln\frac{1+\del}{\del}-1\right),
\\
C^\hi_0&=&\frac{\lambda\omega}{\del}.
\eeqa
Starting from these values, a monotonic rise with increasing
$g$ is observed for all these quantities.
This will be seen more clearly in the next subsection.

\subsection{Mean-field phase diagram}
\label{mftdiag}

\begin{figure}[!ht]
\begin{center}
\includegraphics[angle=-90,width=.75\linewidth]{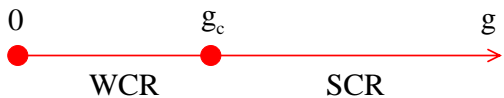}
\caption{\small
(Color online)
Mean-field phase diagram of the model
in the situation where the damping rate $\del$ is very small.
WCR: weak-coupling regime ($g<g_c$).
SCR: strong-coupling regime ($g>g_c$).}
\label{phase}
\end{center}
\end{figure}

The situation of most interest is where the damping rate~$\del$ is small.
In this regime, even in the absence of interactions,
the model already exhibits a broad spectrum of relaxation times
$\tau_i$ (see~(\ref{taui})),
with the largest relaxation time, corresponding to perfectly fit papers ($\eps=1$),
diverging as $\tau_\max=1/\del$.
In the presence of interactions, for $\del\ll1$
the mean-field solution~(\ref{dmf})--(\ref{cmf}) yields a non-trivial
phase diagram (Figure~\ref{phase}),
where a weak-coupling regime (WCR) and a strong-coupling regime (SCR)
are separated by a sharp crossover near the critical coupling
\beq
g_c=\frac{2\del}{\omega}.
\label{gc}
\eeq

From a quantitative viewpoint,
the following predictions can be readily obtained by appropriately expanding
the general mean-field solution~(\ref{dmf})--(\ref{cmf}) in various regimes.

\subsubsection{Weak-coupling regime ($g<g_c$)}

The WCR corresponds to the range of parameters $x$
such that $\e^x$ is comparable with $1/\del$.
In this regime,~(\ref{dmf}) and~(\ref{gmf}) yield after some elementary algebra
\beq
x\approx\ln\frac{1}{(1-g/g_c)\del}.
\eeq
The mean activity $A$ and the highest citation count are then respectively
given by:
\beqa
A&\approx&\omega\left(\ln\frac{1}{(1-g/g_c)\del}-1\right),
\\
C^\hi&\approx&\frac{\lambda\omega}{(1-g/g_c)\del}.
\eeqa
All over the WCR, the damping rate $\del$ is renormalised by the factor
$(1-g/g_c)$,
which vanishes as the critical coupling is approached ($g\to g_c$).

\subsubsection{Strong-coupling regime ($g>g_c$)}

The SCR corresponds to the range of parameters $x$
such that $\e^x$ is much larger than $1/\del$.
In this regime we have
\beq
x\approx\frac{g/g_c-1}{2\del}.
\label{xscr}
\eeq
The prediction for the mean activity $A$ reads
\beq
A\approx A_\infty(1-g_c/g),
\label{as}
\eeq
with
\beq
A_\infty=\frac{1}{g_c}=\frac{\omega}{2\del}.
\label{ainf}
\eeq
We have also
\beq
L\approx M\approx\frac{g}{g_c}.
\label{mflm}
\eeq
These estimates imply that the relaxation rate entering~(\ref{mf2})
vanishes almost linearly with $1-\eps$,
so that very fit papers have very long relaxation times.
The relaxation time of perfectly fit papers ($\eps=1$),
and the corresponding citation count,
\beq
C^\hi\approx\frac{2\lambda\del}{g}\,\e^x,
\label{chst}
\eeq
with $x$ given by~(\ref{xscr}),
are exponentially large in $1/\del$ all over the SCR.

\subsubsection{Critical point ($g=g_c$)}

In the borderline situation where the coupling constant is at its critical value ($g=g_c$),
the critical value $x_c$ of the parameter $x$ satisfies
\beq
2(x_c-1)\e^{x_c}\approx\frac{1}{\del^2}.
\eeq
It therefore diverges essentially logarithmically, as
\beq
x_c\approx\ell-\ln\ell+\frac{1+\ln\ell}{\ell}+\cdots,
\eeq
with
\beq
\ell=\ln\frac{1}{2\del^2}.
\eeq
The prediction for the critical mean activity $A_c$ reads
\beq
A_c\approx\omega(x_c-1),
\eeq
while we have
\beq
C^\hi_c\approx\frac{\lambda\omega}{2(x_c-1)\del^2}.
\eeq

\begin{figure}[!ht]
\begin{center}
\includegraphics[angle=-90,width=.75\linewidth]{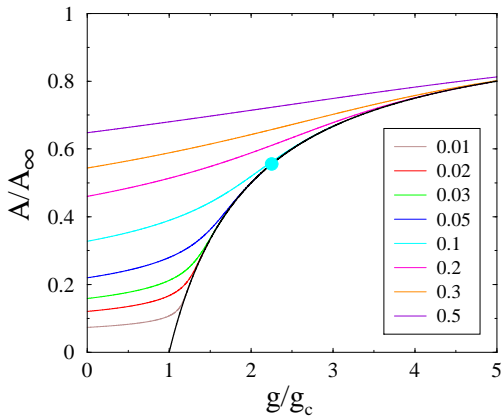}
\caption{\small
(Color online)
Mean-field prediction for the reduced mean activity $A/A_\infty$
against $g/g_c$ for damping rates $\del$ ranging from 0.01 to 0.5 (bottom to top).
Black line: SCR prediction~(\ref{as}).
Symbol: location of the threshold coupling $g_\eff\approx0.45$
(see~(\ref{geff}))
for $\del=0.1$ and $\omega=1$, so that $g_c=0.2$.}
\label{amf}
\end{center}
\end{figure}

Figure~\ref{amf} illustrates the above results.
It shows a plot of the reduced mean activity $A/A_\infty$
against~$g/g_c$ for damping rates $\del$ ranging from 0.01 to 0.5.
The black line shows the SCR prediction~(\ref{as}).
For each $\del$, there is a threshold coupling $g_\eff(\del)$ beyond which
the latter prediction suddenly becomes very accurate.
When $\del$ is very small (lower curves),
the crossover between WCR and SCR is very sharp,
and $g_\eff(\del)$ is very close to the predicted threshold~$g_c$ (see~(\ref{gc})).
For larger values of $\del$ (upper curves), the crossover becomes broader,
whereas $g_\eff(\del)$ progressively becomes larger than~$g_c$.
This incipient discrepancy with increasing $\del$ is to be expected,
since the analytical prediction for $g_c$
was derived in the limit of a very small damping rate $\del$.

\section{Numerical results}
\label{num}

This section comprises extensive numerical explora\-tions of various aspects of our model.
In Section~\ref{global}, we examine the behaviour of a global quantity such as
the total activity.
The fates of individual papers, including a test of the validity
of mean-field theory in this case, are studied in Section~\ref{single}.
Next, in Section~\ref{fittest}, we look at
the sequences of successive papers ranked by high fitness and citation counts.
Finally, we present some statistical information on the distribution
of the highest citation counts (Section~\ref{counts}).

All the results of this section have been obtained by means
of a direct numerical solution of the coupled differential
equations~(\ref{eqeta})
describing the evolution of the dynamic fitnesses $\eta_i(t)$.
Mean values of observables are defined as averages over these fitnesses,
i.e., over the whole stochastic history of the model.
Similarly, probabilities are defined with respect to the ensemble of all such histories.
A major simplification results from the fact that we can use the
expression~(\ref{cmean}),
without having to actually simulate the full stochastic citation process, as we
are only interested in mean citation counts.
Also, as the short-time dynamics of the model are entirely irrelevant,
we can safely replace the random publication times of papers by regularly spaced times.
Unless stated otherwise, we choose $\omega=1$ from now on,
so that paper number $t$ is published at the integer time~$t$, and $\del=0.1$.
For these parameter values, the onset of the SCR (see Figure~\ref{amf}) reads
\beq
g_\eff\approx0.45.
\label{geff}
\eeq
This number, shown as a symbol in Figure~\ref{amf},
is roughly twice the prediction $g_c=0.2$
(see~(\ref{gc})), which holds in the $\del\to0$ limit.

\subsection{Total activity}
\label{global}

\begin{figure}[!ht]
\begin{center}
\includegraphics[angle=-90,width=.75\linewidth]{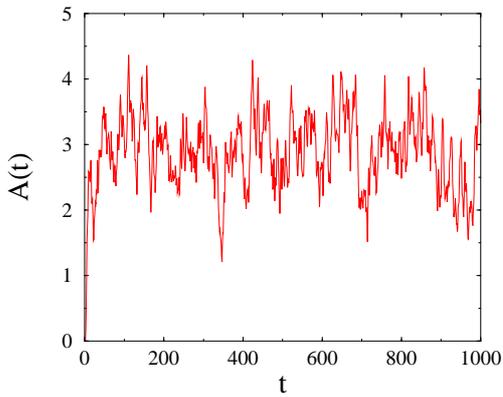}
\caption{\small
(Color online)
Total activity $A(t)$ against time $t$ during a single history of $N=1000$
papers
with $\omega=1$, $\del=0.1$ and $g=0.5$.}
\label{aplot}
\end{center}
\end{figure}

The total activity
\beq
A(t)=\sum_i\eta_i(t),
\eeq
where the sum runs over papers $i$ published before time~$t$,
is the fluctuating counterpart of the mean activity~$A$, discussed above.
It is also the simplest of all global quantities.
Figure~\ref{aplot} shows a plot of $A(t)$
during a single history of $N=1000$ papers with $g=0.5$.
The system soon reaches a steady state,
where the activity keeps fluctuating around a well-defined mean value.

\begin{figure}[!ht]
\begin{center}
\includegraphics[angle=-90,width=.75\linewidth]{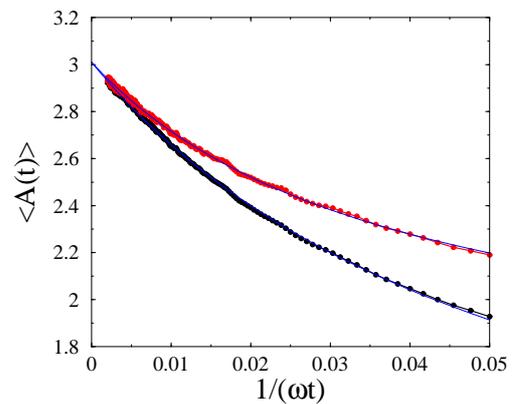}
\caption{\small
(Color online)
Mean total activity $\mean{A(t)}$ against $1/(\omega t)$.
Black (lower) dataset: $g=0.5$ and $\omega=1$.
Red (upper) dataset: $g=1$ and $\omega=0.5$.
Each dataset is averaged over 10,000 independent histories of $N=500$ papers.
Blue lines: two-parameter fits with common intercept 3.01.}
\label{acv}
\end{center}
\end{figure}

There is, however, a subtlety, which is not visible on Figure~\ref{aplot}; this concerns
the very slow relaxation dynamics whereby the steady state is reached, as shown
in Figure~\ref{acv}.
Here, the mean total activity $\mean{A(t)}$ is plotted against $1/(\omega t)$
for two specific situations, $g=0.5$ and $\omega=1$ (lower dataset)
and $g=1$ and $\omega=0.5$ (upper dataset), such that the condition
$g\omega=0.5$ is maintained.
Both datasets converge to the common limit~3.01, as shown by
two-parameter fits (blue lines).
This limit is in excellent agreement
with $A=3.009\,82$, the mean-field prediction~(\ref{acmf}) for the mean activity,
and shows how well mean-field theory works
for steady-state values of such global quantities.
On the other hand, the slow relaxation in $1/(\omega t)$ is quite unusual,
since one would normally expect the steady state to be attained exponentially fast.
Here, this effect can be explained by using
extreme-value statistics~\cite{gumbel,feller,galambos}.
At time $t$, the system only contains a finite number of papers, $n=\omega t$.
In other words, time serves as a measure of the system size.
In particular, at time $t$ the fitness distribution will only have been sampled~$n$ times,
so that the largest initial fitness met up to time $t$ reads
\beq
\eps_\max\approx1-\frac{x}{n},
\label{epsmax}
\eeq
where the random variable $x$ has the exponential probability distribution
$\e^{-x}$.
The fitness distribution is therefore rescaled down by a finite-size correction
of the order of $1/n=1/(\omega t)$.
Hence all global observables are expected
to exhibit slow relaxations in $1/(\omega t)$ to their steady-state values.

\subsection{The fates of individual papers}
\label{single}

The finite-size effects leading to the slow power-law relaxation of global observables
discussed above also turn out to affect individual papers.
These effects are expected to be largest in the SCR and for very fit papers.
That these constitute special cases is already evident from mean-field theory,
where~(\ref{mf2}),~(\ref{mflm}) reveal a very slow decay of their dynamic fitnesses.

\begin{figure}[!ht]
\begin{center}
\includegraphics[angle=-90,width=.75\linewidth]{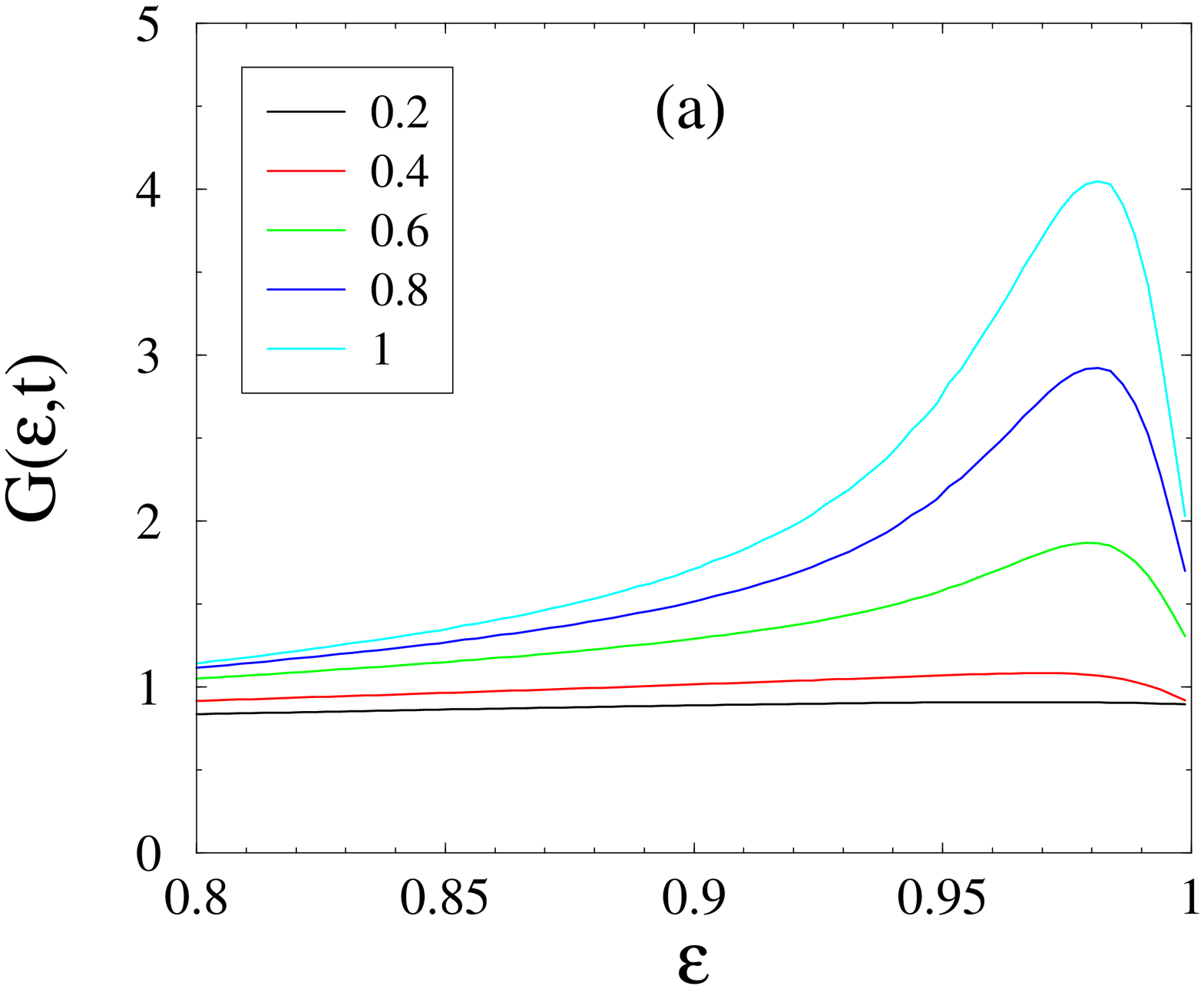}

\includegraphics[angle=-90,width=.75\linewidth]{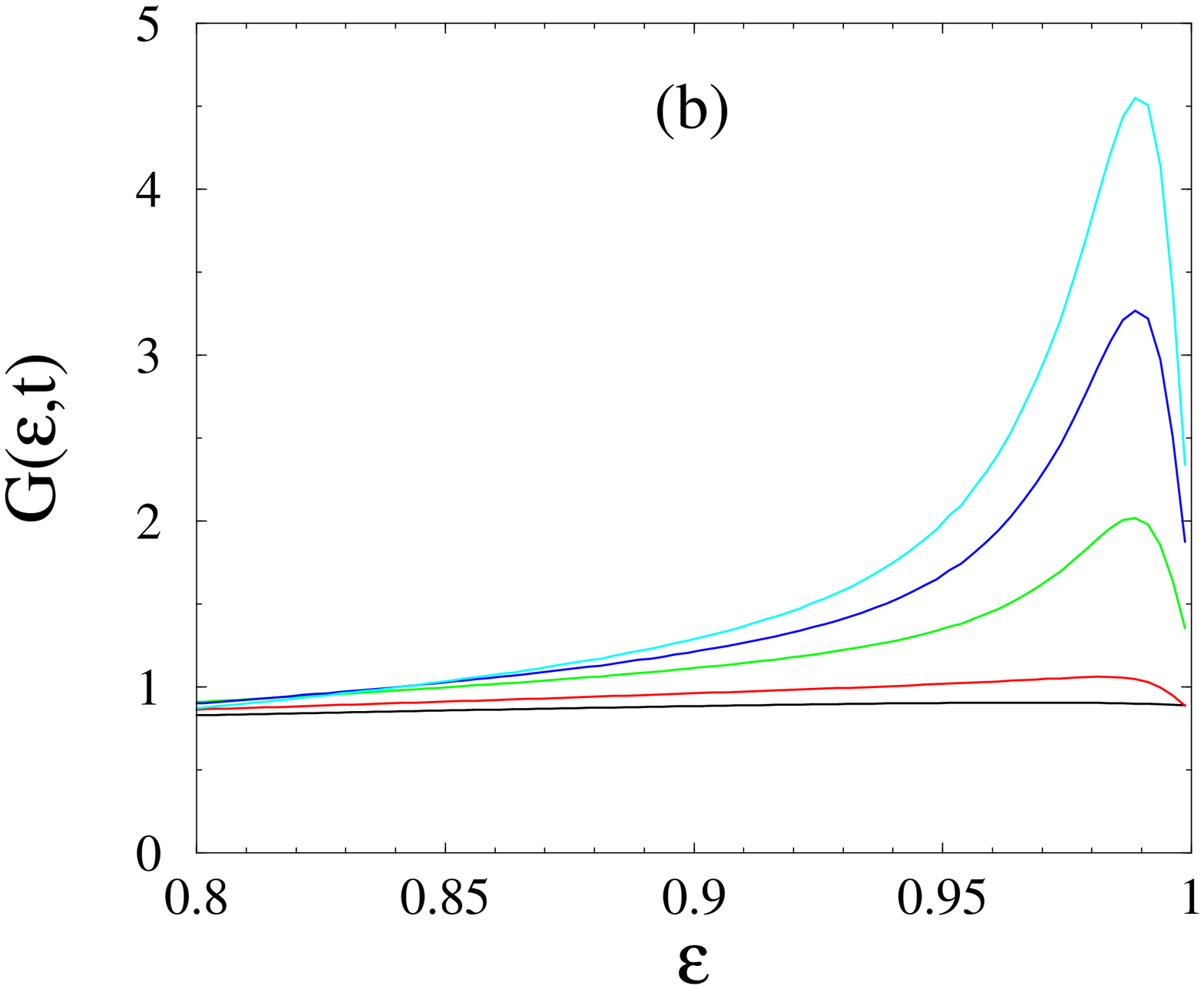}
\caption{\small
(Color online)
Fitness-resolved gain $G(\eps,t)$ against $\eps$,
for (a) $t=100$ and (b) $t=200$,
and values of $g$ ranging from 0.2 to 1 (bottom to top in each panel).}
\label{g}
\end{center}
\end{figure}

We first compare numerical results and analytical predictions for the mean
citation counts of individual papers of given initial fitness $\eps$.
For a given observation time~$t$, the fitness-resolved gain,
\beq
G(\eps,t)=\bigmean{\frac{C_i(t)}{C_\mf(t-i;\eps)}},
\eeq
is defined by averaging the citation counts $C_i(t)$ of all papers
$i=1,\dots,t$
whose initial fitnesses $\eps_i$ are close to~$\eps$.
The mean-field prediction $C_\mf(t-i;\eps)$ in the denominator is given
by~(\ref{mfc}).
Figure~\ref{g} shows histogram plots of $G(\eps,t)$ against $\eps$
for times $t=100$ and $t=200$, and several coupling
constants $g$ denoted by different colours.
In order to focus on the high-fitness end which is the region of most interest,
data are only shown for $\eps>0.8$.

For a fixed suboptimal fitness ($\eps<1$),
the gain $G(\eps,t)$ approaches unity in the long-time limit.
Mean-field theory is thus clearly appropriate to describe the citation counts
of papers in the long-time regime, provided their fitnesses are suboptimal.
On the other hand, all over the SCR, i.e., for $g>g_\eff$ (see~(\ref{geff})),
the gain also exhibits a peak near the upper edge of the fitness distribution.
The height of the gain peak stays roughly constant with increasing time;
its location approaches the upper edge ($\eps\to1$) while its width shrinks to zero.
This is a strong indication that {\it the gain peak is due to the few fittest papers}
of a typical history.
We conclude that, while suboptimal papers are well described by mean-field theory,
the fittest papers in the SCR need more sophisticated treatment.

\subsection{The fates of exceptional papers}
\label{fittest}

The results of the previous section lead us to a different
way of examining exceptional papers -- i.e., those that are the fittest and/or
the most cited.
In this approach, we are inspired by a body of literature
on growing networks~\cite{ba,doro,bb1,bb2,kr,bogu,glrec,ggl,gl2}, in the
context of which we refer to the fittest papers as `records',
and the most cited papers as `leaders'.

\smallskip
\noindent {\it Fittest papers (records).}
The fittest paper $I_t$
is the paper with the largest initial fitness encountered until time $t$:
\beq
\eps_{I_t}=\eps_\max(t)=\max(\eps_1,\dots,\eps_t).
\eeq
This continues to be the fittest paper until a paper with larger initial
fitness is published.
A sequence of such papers, each one adjudged the fittest at its
time, can be characterised as a sequence of records, whose statistics
have been widely studied~\cite{ren,gli,rec1,rec2}.
A key result in this field is that the `record-breaking probability',
i.e., the probability that paper $t$ is the fittest published so far,
is nothing but $1/t$.
The mean number $N_t$ of such papers up to time~$t$ thus grows logarithmically, as
\beq
N_t=\sum_{i=1}^t\frac{1}{i}\approx\ln t+\gamma,
\label{aver}
\eeq
where $\gamma\approx0.577\,215$ is Euler's constant.

\smallskip
\noindent {\it Most cited papers (leaders).}
The most cited paper $J_t$ at time $t$ has the highest citation count:
\beq
C_{J_t}(t)=C_\max(t)=\max\bigl(C_1(t),\dots,C_t(t)\bigr).
\eeq
This too maintains its position until it is superseded by a newer paper
with a higher citation count.
A sequence of such papers, each with the highest
citations at a given time, is known as a sequence of leaders;
we denote the mean number of such leaders up to time $t$ by $L_t$.
In the present model, a former leader cannot come back to the lead again,
so that the mean number of lead changes up to time~$t$ is $L_t-1$.

In the absence of interactions ($g=0$),
the fittest papers (records) usually become the most cited papers (leaders)
as the following simple argument shows.
For $t$ much larger than the microscopic time scale $1/\del$ fixed by the regulator,
the mean citation count of a paper is given by~(\ref{mfcinf}),
with $L=1+\del$ and $M=1$ (see~(\ref{ablm})), i.e.,
\beq
C_i\approx\frac{\lambda\omega\eps_i}{1+\del-\eps_i}.
\label{cfree}
\eeq
This expression is an increasing function of $\eps_i$ which indicates that,
except in a brief transient regime, the most citations (highest $C_i$) indeed
go to the fittest papers (highest $\eps_i$).
As mentioned above,
the sequences of records and leaders are thus essentially identical;
in particular we predict a logarithmic growth law for the mean number of
leaders at $g=0$:
\beq
L_t\approx N_t\approx\ln t.
\label{lnz}
\eeq

\begin{figure}[!ht]
\begin{center}
\includegraphics[angle=-90,width=.75\linewidth]{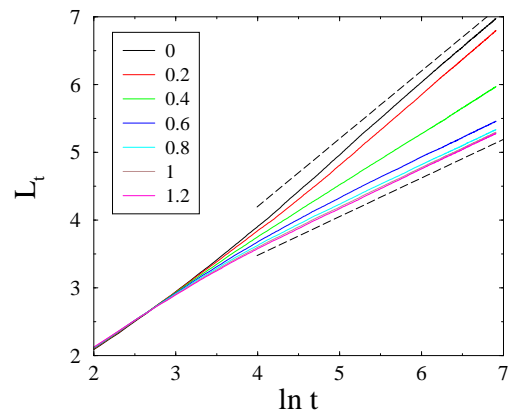}
\caption{\small
(Color online)
Mean number $L_t$ of successive most cited papers (leaders) up to time $t$
against $\ln t$ for coupling constants $g$ ranging from 0 to 1.2 (top to
bottom).
The dashed lines have slopes $C_0=1$ and $C_\scr=0.57$.}
\label{l}
\end{center}
\end{figure}

Figure~\ref{l} shows a plot of the mean number $L_t$ of leaders up to time $t$
against $\ln t$ for coupling constants $g$ ranging from 0 to 1.2.
The data suggest that a logarithmic growth of the form
\beq
L_t\approx C\ln t
\label{avel}
\eeq
holds for all values of the coupling constant $g$.
Another interesting feature is that the amplitude $C$ exhibits a rather sharp
crossover around $g\approx g_\eff\approx0.4$ (green track),
from the value $C_0=1$ in the WCR (in agreement with~(\ref{lnz}) for $g=0$),
to a non-trivial asymptotic value $C_\scr\approx0.57$ deep in the SCR.
We will put these results in perspective with other models in the literature in
Section~\ref{disc}.

In order to explore the statistical properties of records and leaders further,
we define the following two probabilities:
\beq
P_t=\prob\left(J_t=I_t\right)
\label{pdef}
\eeq
is the probability that the leader (most cited paper) is the record (current
fittest paper) at time $t$, while
\beq
\Pi_t=\prob\left(J_t\in\{I_1,\dots,I_t\}\right)
\label{pidef}
\eeq
is the probability that the leader at time $t$ belongs to the sequence of records.
We recall that probabilities are defined with respect to the ensemble of all
stochastic histories of the model.

\begin{figure}[!ht]
\begin{center}
\includegraphics[angle=-90,width=.75\linewidth]{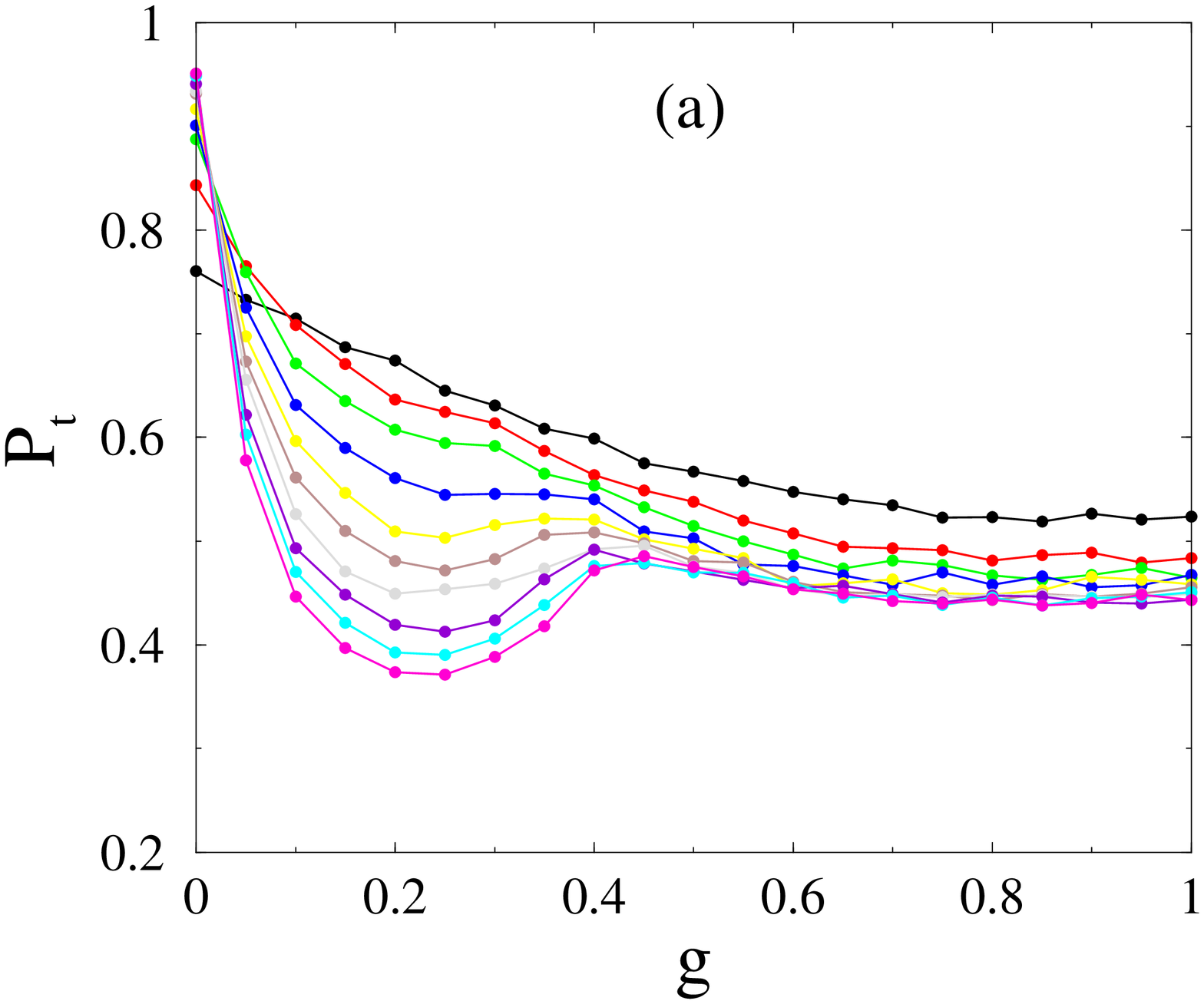}
\includegraphics[angle=-90,width=.75\linewidth]{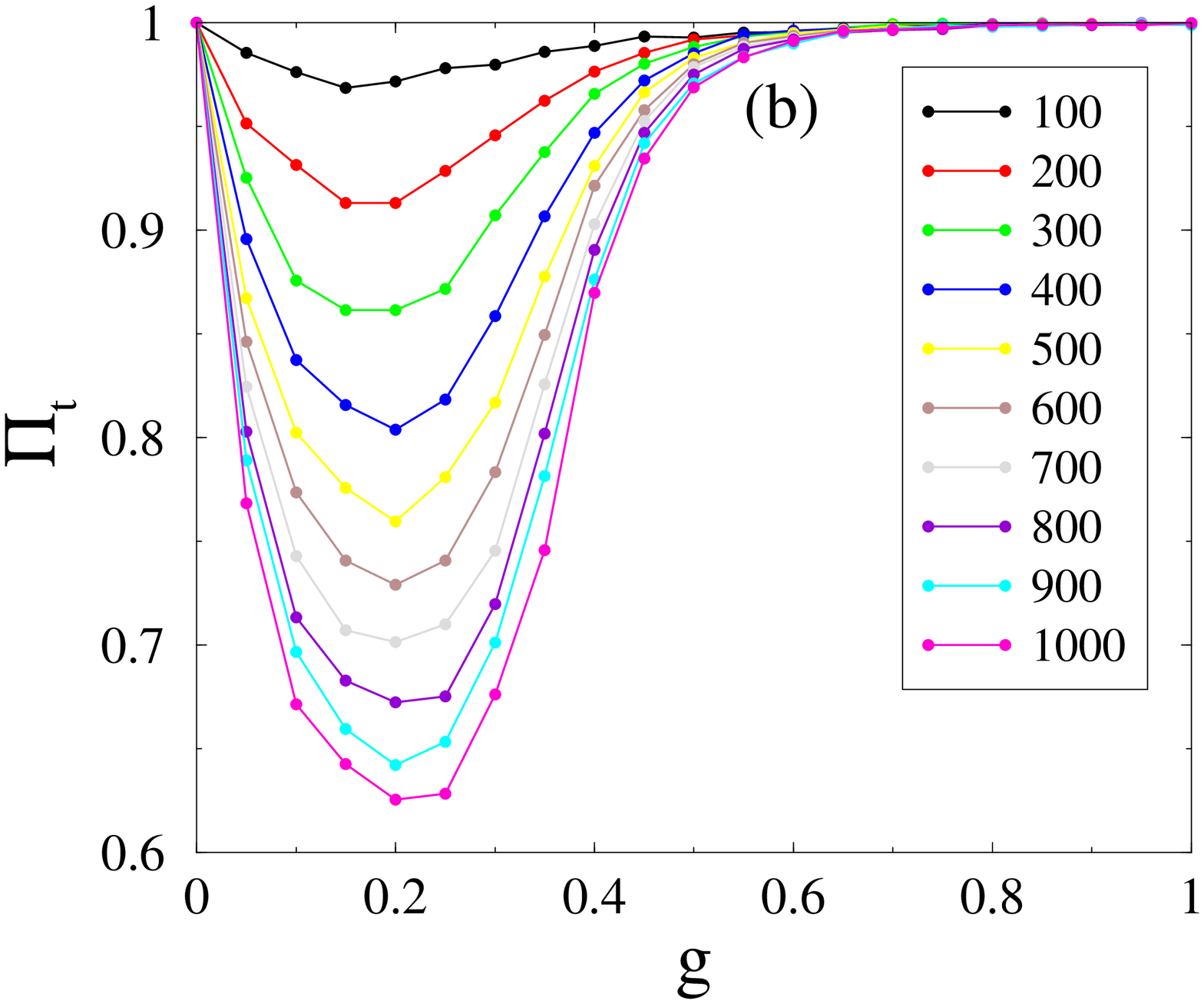}
\caption{\small
(Color online)
(a) probability $P_t$ that the most cited paper at time~$t$ is the current
fittest paper;
(b) probability $\Pi_t$ that the most cited paper at time~$t$
belongs to the sequence of fittest papers.
Both probabilities are plotted against $g$ for several times $t$
ranging from 100 to 1000 (top to bottom in each panel).}
\label{pg}
\end{center}
\end{figure}

These probabilities are plotted in Figure~\ref{pg} against $g$ for several
fixed times $t$.
In the absence of interactions ($g=0$),
we have $P_t\approx\Pi_t\approx1$ for large times,
in agreement with the above observation that
the sequences of records and leaders are essentially identical.
Beyond this, however, one sees a dramatic dependence on~$g$ in both probabilities,
with strongly different behaviour in the weak- and strong-coupling regimes.

In the WCR, which is effectively defined by $g<g_\eff$ (see~(\ref{geff})),
both probabilities exhibit marked minima
near the middle of the WCR ($g\approx g_\eff/2\approx0.2$).
The minima of $\Pi_t$ are more symmetric and more pronounced.
The observed slow decay of both minima with time
suggests that both probabilities $P_t$ and $\Pi_t$ converge to zero in the whole WCR,
albeit logarithmically slowly.
As a consequence, the most cited papers are in general not among the fittest ones,
at least for very late times.

In the SCR ($g>g_\eff$), the probability $\Pi_t$ is very close to unity,
implying that the most cited paper is almost certainly a record,
i.e., one of the successive fittest papers.
The probability $P_t$, on the other hand, seems to converge
to a non-trivial asymptotic value $P_\scr\approx0.44$ deep in the SCR.
A very weak residual dependence of this asymptotic value on $g$
cannot, however, be ruled out.
The above results suggest that
the theory of records might be an appropriate way of further investigating
the dynamics of the model in its most interesting regime, i.e., deep in the SCR.
While we will return to these considerations in Section~\ref{effective}, it is
well worth re-emphasising the strikingly counter-intuitive
results obtained above: leaders are almost always {\it not} records in the
WCR for asymptotic times, and even in the SCR where a leader is in general one
of the records, {\it it is not necessarily the fittest among them}.

\subsection{The statistics of highest citation counts}
\label{counts}

In this section, we complement the above analysis
by investigating the statistics of the highest citation counts.
These, too, show qualitatively different behaviour in the WCR and SCR.
We have chosen to monitor two observables
which are selectively sensitive to high citation counts.
The first one is self-explanatory:
it is the largest mean citation count $C_\max(t)$ at time $t$.
The second observable is the so-called moment ratio, defined as
\beq
Y(t)=\frac{\sum_i C_i(t)^2}{\bigl(\sum_iC_i(t)\bigr)^2},
\label{ydef}
\eeq
where the sums run over papers $i$ published before time~$t$,
Such dimensionless moment ratios have been widely used to investigate
classical disordered systems;
Derrida and Flyvbjerg~\cite{df} used them to investigate
the statistics of random objects such as valleys in spin glasses as well as in
models of fragmentation, and they have been widely used since to examine other complex
systems~\cite{dp,drev,kgb,bebo,bbp,blm}.
A similar quantity known as the inverse participation ratio
(IPR)~\cite{bdh,bd,vis}
is widely used as a measure of the spatial extension of wavefunctions in quantum systems.
In the context of Anderson localisation, its use allows one to
distinguish between extended and localised states~\cite{th,fifty}.

\begin{figure}[!ht]
\begin{center}
\includegraphics[angle=-90,width=.75\linewidth]{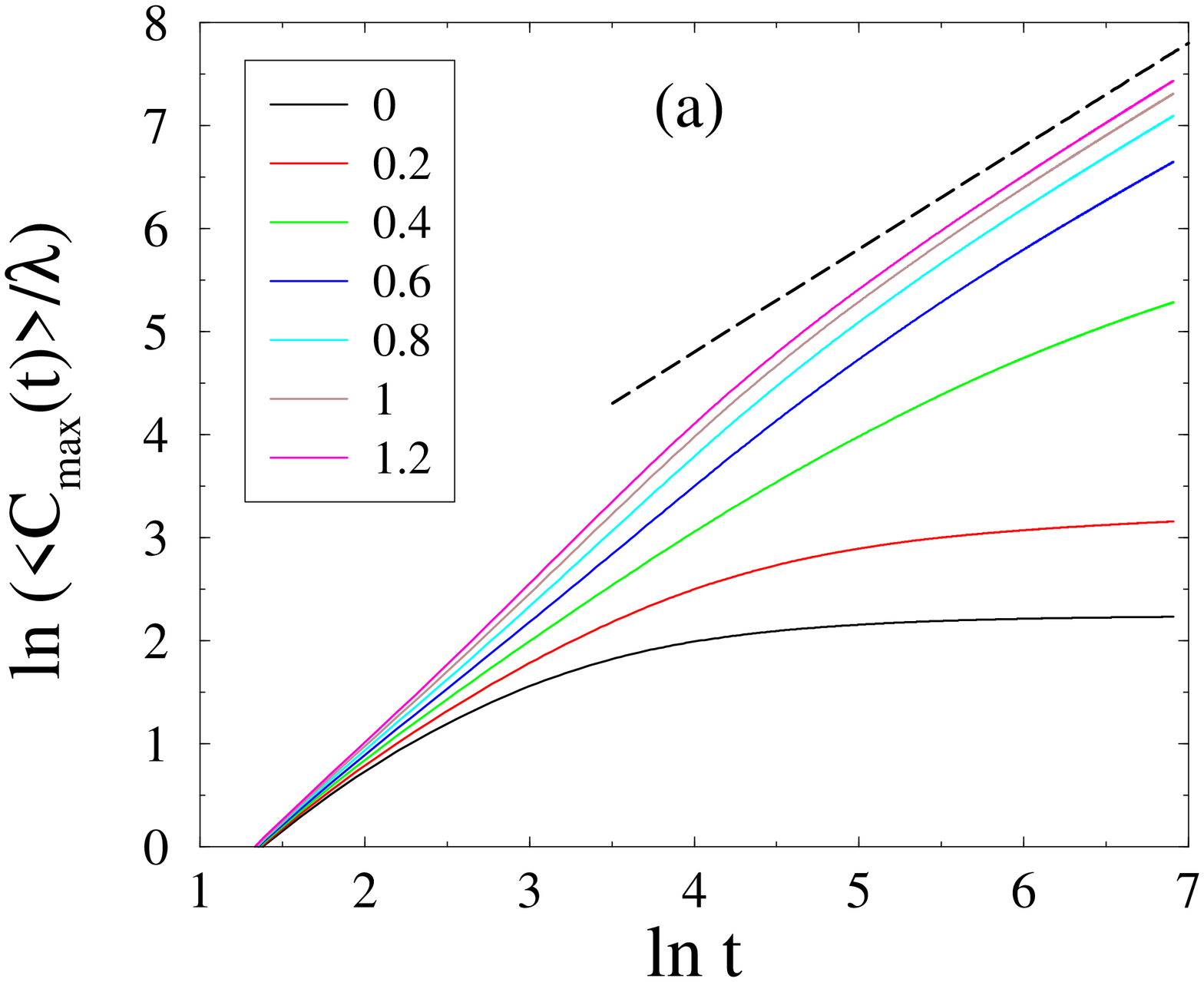}

\includegraphics[angle=-90,width=.75\linewidth]{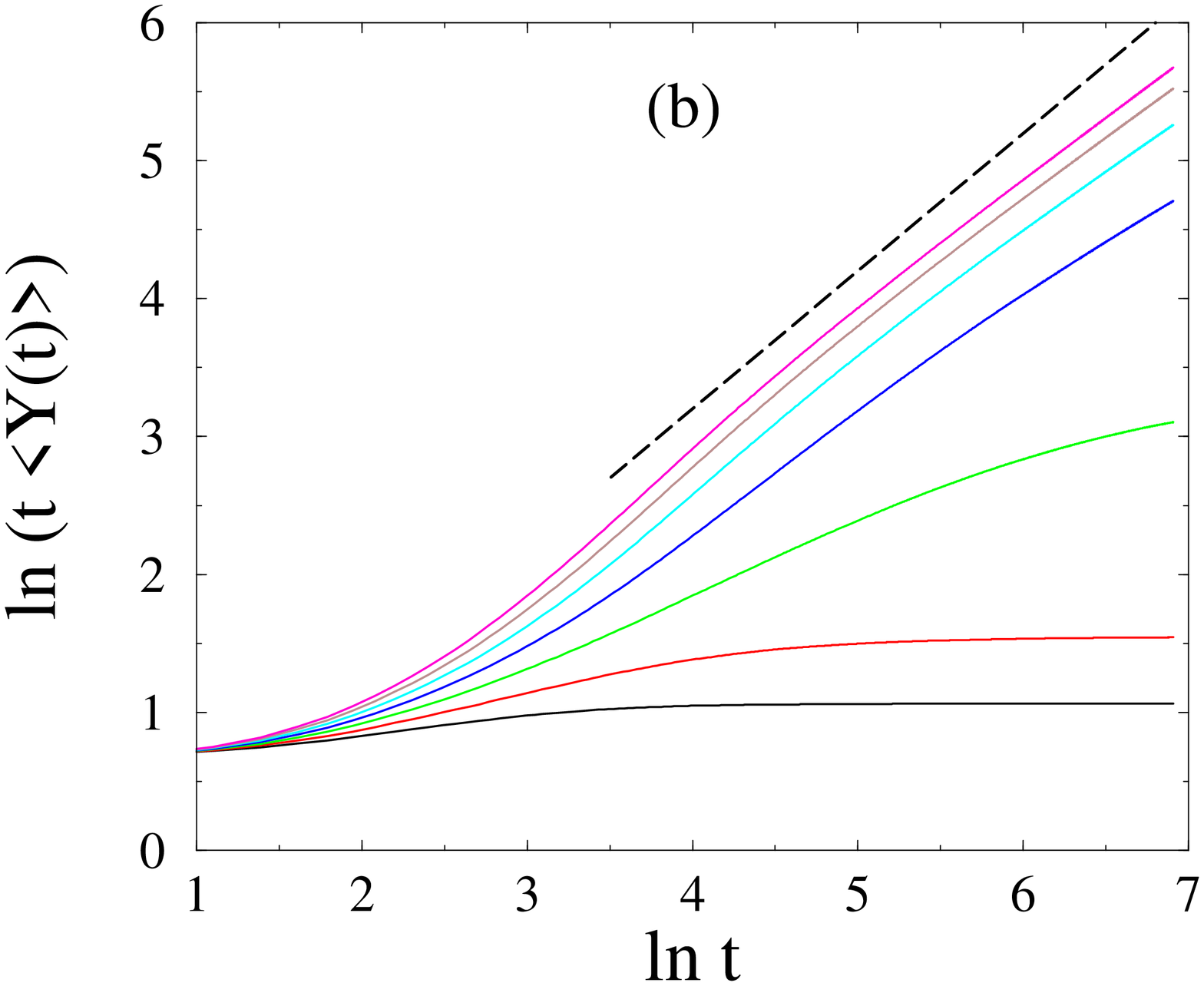}
\caption{\small
(Color online)
Log-log plots of (a) the average $\mean{C_\max(t)}$ of the largest citation count
in units of $\lambda$
and (b) the product $t\mean{Y(t)}$, against $t$,
for coupling constants $g$ ranging from 0 to 1.2 (bottom to top in each panel).
Dashed lines with unit slope are guides to the~eye.}
\label{cy}
\end{center}
\end{figure}

Here, the statistics of the quantity $Y(t)$ will be used to highlight the difference
between the relatively featureless, `extended' distribution of citation counts
in the WCR and the strongly rugged, `localised' distribution in the SCR,
where by contrast a few papers dominate the overall distribution
with their huge citation counts.
Figure~\ref{cy} shows log-log plots
of the average of the largest citation count at time $t$, $\mean{C_\max(t)}$,
and of the product $t\mean{Y(t)}$, against $t$; the
values of the coupling constant $g$ are chosen to be the same as in Figure~\ref{l}.
Both quantities again exhibit a crossover
around $g\approx g_\eff\approx0.4$ (green tracks).

In the WCR, the largest citation count saturates to a finite value,
or possibly grows very slowly in time,
while the moment ratio $Y(t)$ falls off essentially as $1/t$.
Both indicators point toward a rather flat and structureless distribution
of citation counts among many papers, with relatively few fluctuations.
This would correspond to
an `extended' regime, in the language of Anderson localisation.

\begin{figure}[!ht]
\begin{center}
\includegraphics[angle=-90,width=.75\linewidth]{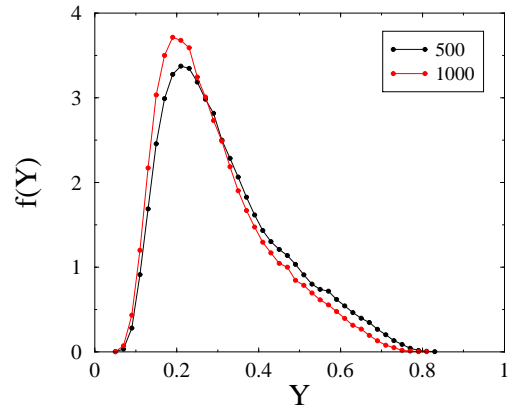}
\caption{\small
(Color online)
Probability distribution of the moment ratio $Y$ for $g=1.2$
and $t=500$ and $t=1000$.}
\label{yhisto}
\end{center}
\end{figure}

The picture in the SCR is entirely different, though; here,
the maximal citation count grows approximately linearly in time,
and the mean moment ratio $Y(t)$ slowly converges to a non-trivial limit.
This is a clear signature that strong fluctuations persist
even in the `thermodynamic' limit of very long times.
These observations are corroborated by Figure~\ref{yhisto} which
shows a histogram plot of the probability distribution of $Y$
for $g=1.2$ (which is deep in the SCR) for two large times, $t=500$ and $t=1000$.
Despite the undoubted presence of finite-size effects, there is a noticeable
convergence towards a non-trivial asymptotic distribution $f(Y)$,
demonstrating that fluctuations are neither small nor trivial.
This limit distribution is observed to be very asymmetric,
and vanishes exponentially fast at both endpoints ($Y\to0$ and $Y\to1$).

Such distributions of the moment ratio $Y$
are characteristic of {\it non-self-averaging} systems~\cite{df,drev},
where strong history-dependent fluctuations ensure
that even very large systems cannot incorporate all possible fluctuations.
As a consequence, observables fail to self-average in the thermodynamic limit.
In our case, this comes about firstly because of the emergence of a strongly
hierarchical distribution of high individual citation counts,
such that the largest counts are finite fractions of the total sum; and secondly,
because these largest counts fluctuate strongly between different histories.
A somewhat similar phenomenon has been observed in a model for the dynamics
of movie competition~\cite{rev5}:
the late-time competition there observed between the best movies, characterised
by very slow oscillations in their popularities,
would yield a similar distribution $f(Y)$ to that of Figure~\ref{yhisto}.
Conversely,
we would also expect to see such slow oscillations between the dynamic fitnesses
of the fittest papers in our model, in a given stochastic history.

All the results of this section underline the special role played by
exceptional papers -- leaders and records, especially in the SCR.
In the next section, we present an effective model
of the citation counts of such exceptional papers deep in the SCR.

\section{Effective model deep in the strong-coupling regime}
\label{effective}

\subsection{Construction}
\label{recursive}

We propose here an effective model of the main features of exceptional papers,
i.e., those with large fitnesses and large citation counts, deep in the SCR.
Our model is based on the following observations:

\begin{enumerate}

\item
Deep in the SCR, the most cited paper is almost certainly one of the records
(successive fittest papers).
This is manifested by $\Pi_t\approx1$,
as can be seen from Figure~\ref{pg}(b).
This suggests that we restrict the dynamics to the sequence of records.

\item
Deep in the SCR, the mean-field prediction $L\approx M$ (see~(\ref{mflm}))
implies that the relaxation rate entering~(\ref{mf2})
vanishes essentially linearly with $1-\eps$,
so that very fit papers have very long relaxation times.
This will be used as a prescription to model the dynamics of records.

\end{enumerate}

We therefore set $\omega=1$ as above, and $L=M=\Omega$,
keeping the effective rate~$\Omega$ as a phenomenological, and in fact the only,
parameter of our effective model.

Within this framework, a very fit paper published at time $t_0$
with initial fitness $\eps$ has a dynamic fitness
\beq
\eta(t)=\e^{-\Omega(1-\eps)(t-t_0)}
\label{etarec}
\eeq
and a mean citation count
\beq
C(t)=\frac{\lambda}{\Omega(1-\eps)}\Bigl(1-\e^{-\Omega(1-\eps)(t-t_0)}\Bigr).
\label{crec}
\eeq

The actual construction of the effective model is based on
record statistics~\cite{ren,gli,rec1,rec2}.
Consider a fixed, very large observation time~$t$, from which
the sequence of successive fittest papers (records) is read backwards.
Using a continuous time formalism,
the fittest paper to date was published at some time $t_1$,
uniformly distributed between~0 and~$t$,
with initial fitness $\eps_1=1-x_1/t$,
where~$x_1$ is drawn from the exponential distribution $\e^{-x_1}$
(see~(\ref{epsmax})).
Similarly, the fittest paper up to time~$t_1$ was published at some time $t_2$,
uniformly distributed between 0 and $t_1$,
with initial fitness $\eps_2=1-x_1/t-x_2/t_1$, such that
$x_2$ has the exponential distribution $\e^{-x_2}$, and so on.
We thus obtain the following recursive scheme.
The publication dates $t_1,t_2,\dots$ of the successive fittest papers,
numbered backwards from time~$t$,
and the corresponding initial fitnesses $\eps_1,\eps_2,\dots$ read
\beq
t_k=t\,s_k,\qquad\eps_k=1-\frac{r_k}{t},
\label{tek}
\eeq
where the dimensionless reduced times $s_k$
and reduced fitnesses $r_k$ obey the random recursions
\beq
s_k=u_ks_{k-1},\qquad r_k=r_{k-1}+\frac{x_k}{s_{k-1}},
\label{srk}
\eeq
with $s_0=1$, $r_0=0$, and therefore
\beqa
s_k&=&u_1\cdots u_k,
\label{skexp}
\\
r_k&=&x_1+\frac{x_2}{u_1}+\frac{x_3}{u_1u_2}+\cdots+\frac{x_k}{u_1\cdots
u_{k-1}}.
\label{rkexp}
\eeqa
The $u_k$ are uniform random variables between 0 and 1,
whereas the $x_k$ are drawn from the exponential probability
distribution~$\e^{-x}$.

The dynamic fitnesses $\eta_k(t)$ and citation counts $C_k(t)$
of the successive records at the observation time $t$
are obtained by inserting the expressions~(\ref{tek})
into~(\ref{etarec}) and~(\ref{crec}).
We thus obtain
\beqa
\eta_k(t)&=&\e^{-\Omega r_k(1-s_k)},
\\
C_k(t)&=&\frac{\lambda t}{\Omega r_k}\left(1-\e^{-\Omega r_k(1-s_k)}\right).
\label{retac}
\eeqa

\begin{figure}[!ht]
\begin{center}
\includegraphics[angle=-90,width=.75\linewidth]{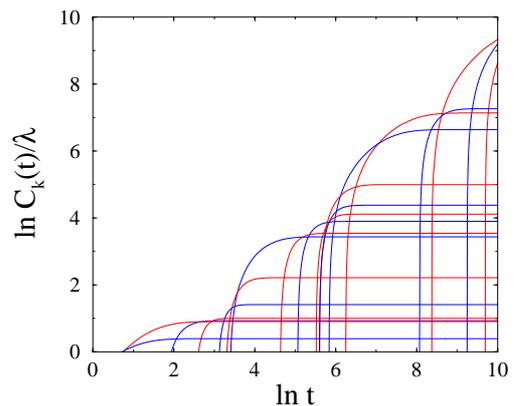}
\caption{\small
(Color online)
Log-log plot of the citation counts $C_k(t)$ in units of $\lambda$
for a single history of the effective model with $\Omega=1$,
against time $t$ up to the observation time $t=\e^{10}\approx22\,026$.
Colors alternate for a clearer reading.}
\label{cplot}
\end{center}
\end{figure}

Figure~\ref{cplot} shows a log-log plot of the citation counts obtained
in units of $\lambda$ for a single history of the effective model with $\Omega=1$,
against time $t$ up to the observation time $t=\e^{10}\approx22\,026$.
Note the regularity of this typical pattern on a logarithmic time scale,
where citation counts rise very fast and level out rather suddenly,
so that every record paper is soon overtaken by a later one,
which in turn is overtaken by one of {\it its} successors, and so on.

An interesting feature of the effective model is its {\it exact
self-similarity.}
The dynamical quantities $\eta_k(t)$ and~$C_k(t)$
only depend on the dimensionless random quantities $s_k$ and $r_k$
(up to an overall factor of $t$ in the citation counts $C_k(t)$).
In the full model, by contrast, the scaling laws which characterise the SCR
(such as the linear growth of $\mean{C_\max(t)}$)
only hold asymptotically for very large times.

The above self-similarity
relies on the choice of a uniform distribution of initial fitnesses.
For a different fitness distribution, e.g.~one which obeys the power law~(\ref{rhop}),
the estimate~(\ref{tek}) would read $1-\eps_k\sim t^{-1/(1-\beta)}$,
so that the dynamic fitnesses $\eta_k(t)$ would acquire
an explicit time dependence, thus breaking scale invariance.
Even for the uniform fitness distribution we consider here,
the self-similarity breaks down at an exponentially large time scale beyond
which our model should not be pushed:
\beq
t^\star\sim\exp\left(\frac{g/g_c-1}{2\del}\right).
\eeq
This is where the citation count $C_1(t^\star)$, say, becomes comparable
to the mean-field estimate~(\ref{chst}) for $C^\hi$.

Before we present the main results of the effective model,
it is worth looking at the fates of very early records, labelled by large $k$,
which were born much before the observation time $t$.
We observe that
\beq
z_k=-\ln s_k=\ln\frac{t}{t_k}
\eeq
is the sum of $k$ positive random variables $y_k=-\ln u_k$,
drawn from the exponential probability distribution $\e^{-y}$
(see~(\ref{skexp})).
The distribution of $z_k$ is therefore a `gamma' distribution of the form:
\beq
f_k(z)=\frac{z^{k-1}\e^{-z}}{(k-1)!}.
\label{fz}
\eeq
We have in particular $\mean{z_k}=k$.
For large $k$, the distribution of $s_k$ is strongly peaked around
the typical value $s_k^\ty=\e^{-\mean{z_k}}=\e^{-k}$.
Putting all this together, we see that the publication times of early records,
as well as their citation counts, typically fall off exponentially with~$k$:
\beq
t_k^\ty\sim C_k^\ty(t)\sim\e^{-k}\,t.
\label{eexp}
\eeq
These simple estimates have important consequences.
First, for a large but finite observation time $t$,
the sequence of fittest papers contains only $k\approx\ln t$ papers --
this estimate being obtained by setting $t_k^\ty\sim1$.
We thus recover the logarithmic law~(\ref{aver})
for the mean number of fittest papers (records),
including its unit prefactor.
Second, even for an infinite history,
the exponential decay of $C_k^\ty(t)$ predicted in~(\ref{eexp})
implies that the number of papers with significant citation counts
remains effectively finite.
These findings are in agreement
with the results of our numerical simulations of the full model
presented in Section~\ref{num}.

\subsection{Main results}
\label{predictions}

Since the effective model is still too complicated to be solved
analytically, we take recourse to numerical simulations for further investigations.
In order to allow for a comparison with the results on the full model
presented in Section~\ref{num},
we focus our attention on the probability~$p_k$ that the $k$th fittest paper
(numbered backwards from the observation time $t$) is the most cited.
The first of these, $p_1$, is the probability that the most cited paper
is the current fittest one.
It is therefore
the analogue of the probability $P_t$ defined in~(\ref{pdef}) for the full
model.
We also investigate the statistics of the moment ratio
\beq
Y=\frac{\sum_k C_k(t)^2}{\bigl(\sum_kC_k(t)\bigr)^2},
\label{rydef}
\eeq
where the citation counts $C_k(t)$ are given by~(\ref{retac}).
The quantity $Y$ thus defined is independent of the observation time $t$,
as a result of the exact self-similarity of the effective model.

\begin{figure}[!ht]
\begin{center}
\includegraphics[angle=-90,width=.75\linewidth]{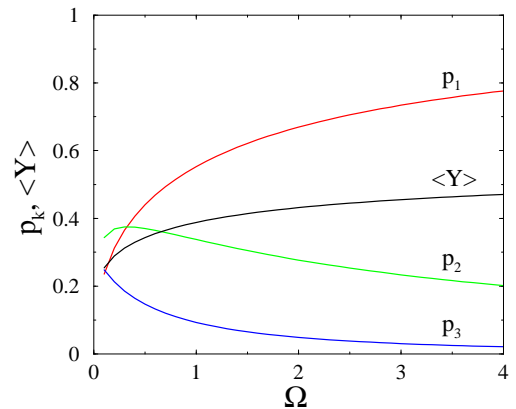}
\caption{\small
(Color online)
The first three probabilities $p_k$ and the mean moment ratio $\mean{Y}$
of the effective model, against the effective rate $\Omega$.}
\label{rpky}
\end{center}
\end{figure}

Figure~\ref{rpky} shows a plot of the first three probabilities~$p_k$
($k=1$, 2, 3) and of the mean moment ratio $\mean{Y}$
against the effective rate~$\Omega$.
The plotted quantities are observed to depend smoothly on $\Omega$.
As $\Omega$ increases, the probability $p_1$ increases steadily,
whereas the other probabilities slowly fall off to zero,
and the mean moment ratio $\mean{Y}$ increases slowly.
It will be seen in Figure~\ref{ryh} that the full distribution of $Y$
also shifts to the right with increasing $\Omega$.
All these observations suggest that the `localisation' features
mentioned in Section~\ref{counts} become more and more pronounced as $\Omega$ is increased.

\begin{figure}[!ht]
\begin{center}
\includegraphics[angle=-90,width=.75\linewidth]{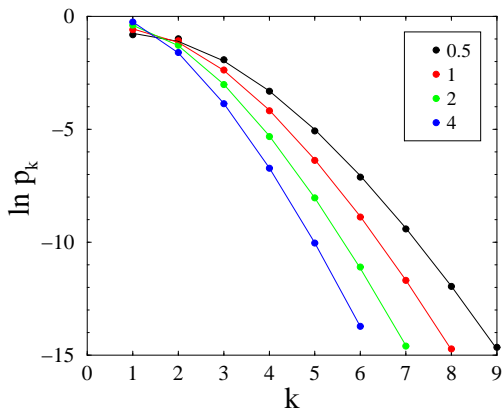}
\caption{\small
(Color online)
Logarithmic plots of the probabilities~$p_k$
against $k$ for four values of the effective rate $\Omega$
ranging from 0.5 to 4 (top to bottom).
Full lines: fits of the form $p_k=A(\mu^k/\Gamma(k+b))^2$,
with $\mu\approx2.48$ for $\Omega=0.5$,
$\mu\approx1.82$ for $\Omega=1$,
$\mu\approx1.26$ for $\Omega=2$,
$\mu\approx0.92$ for $\Omega=4$.
The product $\Omega\mu^2$ is observed to be slightly above 3 in all cases.}
\label{rpklog}
\end{center}
\end{figure}

\begin{figure}[!ht]
\begin{center}
\includegraphics[angle=-90,width=.75\linewidth]{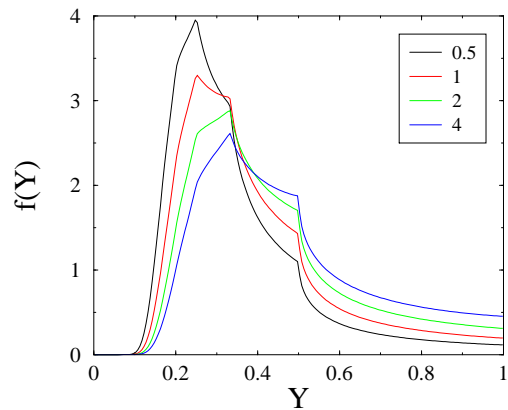}
\caption{\small
(Color online)
Probability distribution $f(Y)$ of the moment ratio~$Y$
for the same values of the effective rate $\Omega$ as in Figure~\ref{rpklog}.}
\label{ryh}
\end{center}
\end{figure}

The effective model allows one to explore more subtle issues,
which cannot be addressed directly in the full model.
One example
concerns the probability that the most cited paper at the observation time $t$
is a very early record, corresponding to a large value of the label~$k$.
Figure~\ref{rpklog} shows logarithmic plots of the probabilities~$p_k$
for a wide range of values of~$k$, and for four values of~$\Omega$.
The smallest of these probabilities is of the order of $10^{-6}$.
Such a figure is far too small to be measurable
by means of a direct numerical simulation of the full model.
The probabilities $p_k$ are clearly observed to decay more rapidly than exponentially.
In the present context, this superexponential behavior can be explained as follows.
Consider a history (i.e., a draw of the random variables~$u_k$ and $x_k$)
such that the most cited paper was published very early on ($k\gg1$).
This history violates the estimates~(\ref{eexp}) quite strongly.
Such atypical behavior can only be obtained if $s_k$ and $r_k$
are of order unity (instead of being exponentially large or small).
Now, for a fixed scale~$\mu$, the probability for having $s_k>\e^{-\mu}$, i.e., $z_k<\mu$,
can be read off from the result~(\ref{fz}): for large $k$, it scales as $\mu^k/k!$.
The constraint on the $r_k$ can be argued
to bring a second factor of the same order of magnitude.
We are thus left with the estimate
\beq
p_k\sim\left(\frac{\mu^k}{k!}\right)^2.
\label{pest}
\eeq
Despite the qualitative nature of the above arguments,
the fits in Figure~\ref{rpklog}
show that the probabilities $p_k$ agree very well
with the superexponential estimate~(\ref{pest}).

The effective model also exhibits another striking feature,
shown in Figure~\ref{ryh}.
Histogram plots of the probability distribution of the moment ratio $Y$
defined in~(\ref{rydef}) are shown for four values of the effective rate~$\Omega$.
The most salient feature of these plots is
the occurrence of singularities at $Y=1/2$, $1/3$, $1/4$, and so on.
Singularities of this kind are ubiquitous in the statistics of random objects
such as attractors in dynamical systems or valleys in disordered
systems~\cite{df,drev}.
Discrete mathematics also contains many instances of distributions with
such singularities in the statistics of
random trees, maps and permutations (see e.g.~\cite{gou}).
In the present situation,
the occurrence of these singularities can be explained in elementary terms.
Consider a history where the $n$ largest citation counts
are almost equal, whereas all other ones are negligibly small.
Such a history yields $Y=1/n+\eps$, where~$\eps$ is very small and positive.
It therefore contributes to $f(Y)$ for $Y=1/n+\eps$, but not for $Y=1/n-\eps$.
For all its roughness, this argument correctly predicts the occurrence
of singularities in $f(Y)$ at all the inverse integers.

Apart from this, the distribution of $Y$ is observed to depend smoothly
and rather weakly on the phenomenological parameter $\Omega$.
The main effect of $\Omega$ is again a shift of $Y$ to larger values for larger $\Omega$.
Furthermore, all over the rather broad range of values of $\Omega$ considered here,
its overall shape reproduces qualitatively the main characteristics
of the distribution observed in the full model deep in the SCR (see Figure~\ref{yhisto}).

In conclusion, our effective model offers at least a qualitative explanation
for the main observed features of the full model deep in the SCR.

\section{Discussion}
\label{disc}

Our main aim in this work has been to examine the competitive dynamics of
high-fitness entities with a view to determine how they become leaders
(or winners/survivors, in the language of earlier models~\cite{us,nirmal}).
The particular paradigm we have used is that of a model citation network,
where individual papers with given initial fitnesses
compete with one another to gain the highest citation counts.
We were inspired to make this choice by findings of universality
in citation statistics~\cite{rev4,wsb},
which we felt might be related to our own findings of universal features
in competitive dynamics on complex networks~\cite{usrec}.
It is however useful to re-emphasise here that
whereas the findings of~\cite{rev4,wsb} related to universality in the citation counts
of individual papers, we have chosen to focus on more collective aspects in our analysis.
Here, the presence
of adaptive dynamics between individual papers results in their fitnesses, and
hence their citation counts, being dynamically modified in the course of time,
as a result of interactions.
Interestingly, our analysis of these more complex collective dynamical quantities
still retains a flavour of universality, of which more will be said below.

Our main result is the emergence of a non-trivial phase diagram,
with a weak-coupling regime (WCR) and a strong-coupling regime (SCR),
separated by a sharp crossover near the critical coupling~$g_c$.
This mean-field picture was corroborated
by an in-depth numerical study of many different facets of the model.
Global observables, such as the total activity, exhibited a slow power-law relaxation
to their steady-state mean values, around which they fluctuated.
The mean-field predictions concerning single papers were found to be
essentially correct,
except in the SCR regime and for very fit papers.
To be specific, there are no interactions between papers at zero coupling,
so that there can be no adaptive dynamics:
the leaders and the records are then essentially identical,
the highest citations going to the papers with greatest initial fitness.
In the WCR, there are few surprises at the mean-field level, where
the mean activity and citation counts increase smoothly as a function of the
coupling constant.
The WCR nevertheless exhibits unexpected behaviour
as correlations between records and leaders are examined:
for large times, it becomes less and less probable that the most cited paper
belongs to the sequence of fittest papers (records).
In the SCR, the fittest papers are shown to have very long relaxation times
and the first signal of non-trivial behaviour is in the fitness-resolved gain,
where a few of papers in the tail of the fitness distribution are seen
to get nearly all the citations.
The investigation of the probabilities $P_t$ and $\Pi_t$
reveals that leaders are certain to belong to the sequence of records,
even if it is not always the fittest among them who win out.

A further probe involving the use of the moment ratio~$Y$
emphasised the qualitative differences between both regimes,
the WCR being characterised by an `extended' and rather structureless
distribution of citation counts among many fit papers,
and the SCR by a `localised' and strongly hierarchical distribution of citation counts,
with only a small number of winners attracting very large citation counts, that
fluctuated strongly between different histories.
This regime was the focus of the effective model of Section~\ref{effective},
which was aimed at capturing the main features
of papers with large fitnesses and large citation counts deep in the SCR.
This self-similar model is based on a recursive construction
of the random sequence of fittest papers (records)
and of the dynamics of their fitnesses and citation counts.
Its results were found to be in qualitative agreement with the numerical
results found earlier.
In particular and importantly, it corroborated the
existence of a non-trivial distribution of the moment ratio,
which is the clearest manifestation of non-self-averaging effects
and of the role of fluctuations in the problem. This distribution provides a focal point
for the investigation of competition among the fittest in future empirical studies
of citation dynamics: a good starting
point could be the plotting of the relative citation counts of leaders
(as done in another context in Ref.~\cite{rev5}), where
the detection of slow oscillations would point towards a detailed examination of the moment ratio, and thus a test of our theory.

We now put our results in perspective with other models, focusing first on
generic growing networks, and next on work specifically related to citation networks.

The record-driven growth process investigated in~\cite{glrec}
models the zero-temperature limit of
a growing network model with preferential attachment
in a rugged fitness landscape
introduced by Bianconi and Barab\'asi~\cite{bb1,bb2}.
The latter model may itself be viewed as an elaboration on earlier models
of complex networks, being either growing~\cite{ba} or static~\cite{rev1,rev2}.
In~\cite{glrec}, a logarithmic law of the form~(\ref{avel}) holds,
with a prefactor $C\approx0.624$,
which is not too far from the prefactor $C_\scr\approx0.57$ found in our work.
However, while the asymptotic probability for the current record
to be the current leader also takes the same value of 0.624 in~\cite{glrec},
our value for this, $P_\scr$ = 0.44, is significantly different,
suggesting that the two models differ significantly in their treatment of correlations.
Additionally we observe that models of growing networks
with preferential attachment of nodes with intrinsic fitnesses,
such as the Bianconi-Barab\'asi model in its low-temperature regime~\cite{gl2},
share the hierarchical features manifested in this work
by the non-trivial histogram of the moment ratio.

Moving on to citation networks,
our work builds on earlier ideas such as the loss of relevance
of papers as they age~\cite{rev3},
on which we base the evolution of our dynamical fitness.
The observation that fitter papers
have longer lifetimes and non-exponential relaxation~\cite{rev3}
is also incorporated in our dynamics.
Empirical studies of citation networks have however usually focused on typical papers,
which are observed to have a broad power-law distribution
of citation counts~\cite{rev3,wsb};
our focus is on the even broader distributions of time scales and citation counts
that are seen for exceptional papers.
In our treatment of those exceptional papers, we find evidence of
extremely nontrivial behaviour which resemble features which have been seen
for exceptional movies~\cite{rev5}.
Last but not least, the importance of ranking has also been underlined~\cite{zz},
which is a reassuring fact in the context of our leaders/records
based approach to exceptional papers.

Finally, and more generally, we suggest that although this model
has been formulated in the context of a citation network,
its results at least in the SCR might be more generally applicable
to problems where a few strongly interacting players
dominate the behaviour of a large assembly,
and where their competitive dynamics result
in huge random fluctuations across different histories.
Although the fight among the fittest could result in leaders
whose identities might fluctuate across histories,
the underlying dynamical processes are strikingly universal.
These processes, involving players in the tail of the fitness distribution,
often lead to the emergence of a leader who, while very fit,
is not actually the fittest of them all.

\begin{acknowledgments}

JML thanks Claude Godr\`eche for interesting discussions
on the statistics of records and related matters.
AM thanks the Department of Bioinformatics, Leipzig, the Institut de Physique
Th\'eorique, Saclay, and the University of Rome `La Sapienza' for their hospitality
during the course of this work.
This project has received funding from the European Research Council (ERC)
under the European Union’s Horizon 2020 Research and Innovation Programme
(grant agreement N.~694925).

\end{acknowledgments}

\appendix

\section{Mean-field theory for an arbitrary fitness distribution}
\label{app}

In this Appendix we extend the mean-field theory of Section~\ref{mft}
to the general situation where the fitness distribution takes an arbitrary form
$\rho(\eps)$ on [0, 1].

The solution of the self-consistency equations~(\ref{asc}),~(\ref{bsc})
for the mean fields $A$ and $B$ now reads
\beqa
A&=&\frac{\omega}{M}\Bigl((1+z)I(z)-1\Bigr),
\\
B&=&\frac{\omega}{M}\Bigl((1+z)^2I(z)-1-z-\ebar\Bigr).
\eeqa
In these expressions, the parameter $z$ reads
\beq
z=\frac{L-M}{M},
\eeq
where $L$ and $M$ are related to $A$ and $B$ by~(\ref{ablm}), whereas
\beq
\ebar=\int_0^1\eps\,\rho(\eps)\,\dd\eps
\eeq
is the mean fitness and
\beq
I(z)=\int_0^1\frac{\rho(\eps)\,\dd\eps}{1+z-\eps}
\eeq
is the Hilbert transform of the fitness distribution.

All the quantities of interest can be expressed in terms
of the parameter $z$ in the range $0<z<\del$,
which satisfies the implicit equation
\beq
(1+z)I(z)=1-\frac{\ebar}{\del-z}+\frac{g\omega\ebar^2}{(\del-z)^2}.
\label{imp}
\eeq
We have in particular
\beq
L=\frac{(1+z)g\omega\ebar}{\del-z},\qquad
M=\frac{g\omega\ebar}{\del-z}.
\eeq
The key quantities of mean-field theory,
namely the mean activity $A$ and the highest citation count $C^\hi$, read
\beqa
A&=&\frac{\omega\ebar}{\del-z}-\frac{1}{g},
\\
C^\hi&=&\frac{\lambda(\del-z)}{g\ebar z}.
\eeqa

In the case of the uniform distribution, considered in the body of this paper,
the Hilbert transform
\beq
I(z)=\ln\frac{1+z}{z}
\label{iuni}
\eeq
is logarithmically divergent as $z\to0$.
The above expression can be parametrized as
\beq
z=\frac{1}{\e^x-1},\qquad I(z)=x,
\eeq
and so we recover the solution given in Section~\ref{mftsol}.

Consider now an arbitrary fitness distribution.
In order for the mean-field solution of the model to be well-behaved for
arbitrary values
of the coupling constant $g$,
the implicit equation~(\ref{imp}) must keep a physically relevant
solution in the range $0<z<\del$ for arbitrarily large~$g$.
The Hilbert transform $I(z)$ therefore has to diverge as $z\to0$.
This condition essentially amounts to saying that the fitness
distribution~$\rho(\eps)$ should not vanish at its upper edge ($\eps\to1$).
In other words, sufficiently many very fit papers should be published at any
time.

The situation where the fitness distribution has a finite density $\rho(1)$
at its upper edge is in every respect similar to the case
of a uniform fitness distribution, studied in detail in Section~\ref{mft}.
In this case, the Hilbert transform indeed diverges as $z\to0$,
albeit only logarithmically:
\beq
I(z)\approx\rho(1)\ln\frac{1}{z}\qquad(z\to0).
\eeq

A qualitatively novel behaviour is observed in the case
where the occurrence of very fit papers is enhanced more significantly,
i.e., where the fitness distribution diverges near its upper edge as a power
law:
\beq
\rho(\eps)\approx K(1-\eps)^{-\beta}\qquad(\eps\to1),
\label{rhop}
\eeq
with an exponent $\beta$ in the range $0<\beta<1$.
Its Hilbert transform also diverges according to the same power law:
\beq
I(z)\approx\KK z^{-\beta}\qquad(z\to0),
\eeq
with
\beq
\KK=\frac{\pi K}{\sin\pi\beta}.
\eeq

Henceforth we focus our attention onto the latter case of a power-law
divergence
of the fitness distribution.
The situation of most interest again corresponds to a small damping rate
($\del\ll1$).
The phase diagram depicted in Figure~\ref{phase} still holds,
with a weak-coupling regime (WCR) and a strong-coupling regime (SCR)
separated by a sharp crossover near the critical coupling
\beq
g_c=\frac{\del}{\omega\ebar}.
\eeq
We obtain the following predictions, which are summarised in Table~\ref{mfexps}.

\begin{table}[!htbp]
\begin{center}
\begin{tabular}{|c|c|c|c|}
\hline
Quantity & $g<g_c$ & $g=g_c$ & $g>g_c$\\
\hline
$z$ & 1 & $2/(\beta+1)$ & $1/\beta$\\
$A$ & $-\beta$ & $-2\beta/(\beta+1)$ & $-1$\\
$C^\hi$ & $-1$ & $-2/(\beta+1)$ & $-1/\beta$\\
\hline
\end{tabular}
\caption{Predictions of mean-field theory for a fitness distribution diverging
at its upper edge as a power law with exponent $\beta$ (see~(\ref{rhop})).
The table gives the exponents governing the power-law behaviour
as $\del\to0$ of the parameter $z$,
the mean activity $A$ and the highest citation count $C^\hi$,
in the three different regimes: WCR ($g<g_c$), critical ($g=g_c$), and SCR
($g>g_c$).}
\label{mfexps}
\end{center}
\end{table}

\begin{itemize}

\item
In the WCR ($g<g_c$), the estimates
\beqa
z&\approx&(1-g/g_c)\del,
\label{zw}
\\
C^\hi&\approx&\frac{\lambda\omega}{(1-g/g_c)\del}
\label{cw}
\eeqa
hold for an arbitrary fitness distribution.

The mean activity $A$ diverges as a power of $\del$:
\beq
A\approx\omega\KK\bigl((1-g/g_c)\del\bigr)^{-\beta}.
\label{awdv}
\eeq

\item
In the SCR ($g>g_c$), the estimates
\beqa
A&\approx&A_\infty(1-g_c/g),
\label{ast}
\\
A_\infty&=&\frac{1}{g_c}=\frac{\omega\ebar}{\del},
\label{ainfst}
\\
L&\approx&M\approx\frac{g}{g_c}
\label{lmst}
\eeqa
hold for an arbitrary fitness distribution.

The parameter $z$ vanishes
and the highest citation count $C^\hi$ diverges as powers of $\del$:
\beqa
z&\approx&\left(\frac{\KK\del}{(g/g_c-1)\ebar}\right)^{1/\beta},
\\
C^\hi&\approx&\frac{\lambda\omega g_c}{g}
\left(\frac{(g/g_c-1)\ebar}{\KK\del}\right)^{1/\beta}.
\eeqa

\item
Right at the critical point ($g=g_c$),
all the relevant quantities of interest obey power laws:
\beqa
z&\approx&\left(\frac{\KK\del^2}{\ebar}\right)^{1/(\beta+1)},
\\
A&\approx&\omega\ebar\left(\frac{\KK}{\ebar\,\del^{2\beta}}\right)^{1/(\beta+1)},
\\
C^\hi&\approx&\lambda\omega\left(\frac{\ebar}{\KK\del^2}\right)^{1/(\beta+1)}.
\eeqa

\end{itemize}

\bibliography{revised.bib}

\end{document}